\documentstyle[12pt,twoside]{article}
{\catcode `\@=11 \global\let\AddToReset=\@addtoreset}
\AddToReset{equation}{section}

\def\greaterthansquiggle{\raise.3ex\hbox{$>$\kern-.75em\lower1ex\hbox{$\sim$}}}
\def\lessthansquiggle{\raise.3ex\hbox{$<$\kern-.75em\lower1ex\hbox{$\sim$}}}

\newcommand{\beq}{\begin{equation}}
\newcommand{\eeq}{\end{equation}}
\newcommand{\beqa}{\begin{eqnarray}}
\newcommand{\eeqa}{\end{eqnarray}}
\newcommand{\beqan}{\begin{eqnarray*}}
\newcommand{\eeqan}{\end{eqnarray*}}
\newcommand{\ba}{\begin{array}}
\newcommand{\ea}{\end{array}}

\newcommand{\dar}{\downarrow}
\newcommand{\uar}{\uparrow}

\newcommand{\Un}{\underline}

\newcommand{\vect}{\overrightarrow}
\newcommand{\ra}{\rightarrow}

\newcommand{\vp}{\varphi}

\newcommand{\A}{{\cal A}}
\newcommand{\B}{{\cal B}}
\newcommand{\C}{{\cal C}}

\newcommand{\Ha}{{\cal H}}

\newcommand{\W}{{\cal W}}

\newcommand{\st}{\stackrel}
\newcommand{\dfrac}{\displaystyle \frac}

\newcommand{\dsum}{\displaystyle \sum}
\newcommand{\dprod}{\displaystyle \prod}

\def\nz{\ifmmode {I\hskip -3pt N} \else {\hbox {$I\hskip -3pt N$}}\fi}
\def\zz{\ifmmode {Z\hskip -4.8pt Z} \else
       {\hbox {$Z\hskip -4.8pt Z$}}\fi}
\def\qz{\ifmmode {Q\hskip -5.0pt\vrule height6.0pt depth 0pt
       \hskip 6pt} \else {\hbox
       {$Q\hskip -5.0pt\vrule height6.0pt depth 0pt\hskip 6pt$}}\fi}
\def\rz{\ifmmode {I\hskip -3pt R} \else {\hbox {$I\hskip -3pt R$}}\fi}
\def\cz{\ifmmode {C\hskip -4.8pt\vrule height5.8pt\hskip 6.3pt} \else
       {\hbox {$C\hskip -4.8pt\vrule height5.8pt\hskip 6.3pt$}}\fi}

\def\au{{\setbox0=\hbox{\lower1.36775ex%
\hbox{''}\kern-.05em}\dp0=.36775ex\hskip0pt\box0}}
\def\lint{\int\limits}
\normalsize
\begin{document}
\bibliographystyle{plain}
\begin{titlepage}
\begin{flushright}
CERN-PH-TH/2005-025\\
February 10, 2005

\end{flushright}

\vspace*{2.7cm}
\begin{center}
{\Large \bf

Supersymmetric Models for Fermions on a Lattice }\\[50pt]

{\sl Nevena Ilieva $^{a,*}$, \,Heide Narnhofer $^{b}$ and
Walter Thirring $^{b}$ }\\[20pt]

{\small
\begin{tabular}{cl}
\qquad & $^a$ Physics Dept., Theory Division, CERN, CH-1211 Geneva
23, Switzerland\\[4pt]
& $^b$ University of Vienna, Institute for Theoretical
Physics\\
& \,\, Boltzmanngasse 5, A-1090 Vienna, Austria, {\sl and}\\[4pt]
& \,\, Erwin Schr\"odinger International
Institute for Mathematical Physics\\
& \,\, Boltzmanngasse 9, A-1090 Vienna, Austria\\[12pt]
\end{tabular}}

\vspace{0.9cm} {\sl We dedicate this article to Julius Wess on the
occasion of his $70^{th}$ birthday.}

\vspace{0.7cm}
\begin{abstract}

We investigate the large-$N$ behaviour of simple examples of
supersymmetric interactions for fermions on a lattice. Witten's
supersymmetric quantum mechanics and the BCS model appear just as
two different aspects of one and the same model. For the BCS
model, supersymmetry is only respected in a coherent superposition
of Bogoliubov states. In this coherent superposition mesoscopic
observables show better stability properties than in a Bogoliubov
state.
\bigskip

\end{abstract}
\end{center}

\vspace{2.6cm}

{\footnotesize

$^\ast$ On leave from Institute for Nuclear Research and Nuclear
Energy, Bulgarian Academy of Sciences, Boul. Tzarigradsko Chaussee
72, 1784 Sofia, Bulgaria

E--mail address: ilieva@ap.univie.ac.at, narnh@ap.univie.ac.at}

\vfill
\end{titlepage}
\setcounter{page}{2}

\section{Introduction}

We realize the simplest supersymmetric system by a finite fermion
lattice. The supersymmetric structure is determined by a
non-hermitian supercharge $Q$. This is an odd nilpotent element of
the Fermi algebra $\A$. The sum and the anticommutator of $Q$ with
its adjoint $Q^\dag $ give two hermitian elements, $G$ and
$H=G^2$. They generate the supertransformation and the time
evolution, two commuting automorphism groups of $\A$. Already at
this level of generality the Hilbert space assumes a structure. It
is the sum of the null-space of these operators plus the tensor
product of $\C^2$ and a rest $\Ha _H.$ In $\Ha _H $ $H$ has
strictly positive eigenvalues, which are twofold degenerate, while
$G$ has the eigenvalues $\pm \sqrt H$ and leaves these spaces
invariant. Thus the supertransformation and the time evolution are
linked. However this closeness is lost once we go to the limit
where $N$, the number of lattice points, goes to infinity. There
it can even happen that on local operators the supertransformation
is not well defined whereas the time evolution is.

To restrict the many possibilities for $Q$, we impose on it a
locality condition: products of operators may only contain
operators of the same lattice site. If there is only one fermion
per site, $G$ and $H$ are essentially unique. The time evolution
becomes trivial, but the supertransformation remains and is in
fact non-local. If there are two fermions per site, the time
evolution becomes the free time evolution and there is a non-local
supertransformation associated to it.

Since even for a local $Q$ the supertransformation is non-local,
we drop the locality requirement. We construct a Clifford variable
$\eta $, which is usually associated with supersymmetry. Note that
a Grassman $\eta$ would mean $\{\eta, \eta^{\dag} \}=0$, which is
not possible in a $\C^*$-algebra, but our $\eta $ is nilpotent and
anticommutes with the other odd elements. With three fermions per
site  we can construct a supersymmmetric version of the BCS model.
This is mathematically well explored and we can behold the many
vistas that the limit $N\rightarrow \infty $ offers. The limit
depends on the state on which the representation is based and we
shall study it for three different states: the ground state of
$H$, its ceiling state, and the Bogoliubov state, which in the
limit $N\rightarrow \infty $    has the same energy per particle
as the ceiling state but remains pure on the quasi-local algebra.

We shall study the following three types of limiting observables:

a) Local operators, i.e. polynomials in the operators $A_i$, the
elements of $\A$, which are localized at the site $i$;

b) Mesoscopic observables, i.e. limits of $\frac{1}{\sqrt{N }}\sum
_{i=1}^N A_i$;

c) Macroscopic observables, i.e. limits of $\frac{1}{N}\sum
_{i=1}^NA_i$.

\smallskip
Our automorphisms turn out to be different in all cases. It is not
even true that the microscopic time evolution determines the
mesoscopic one. The supertransformation is finite and non-trivial
only for the ground state, where Witten's supersymmetric quantum
mechanics emerges in the limiting procedure. It seems remarkable
that the difference between a statistical mixture of the
Bogoliubov states and the ceiling state, which can be considered
as a coherent mixture of the Bogoliubov states, can only be
observed on the mesoscopic and macroscopic level respectively.
Especially the mesoscopic algebra is stable in the ceiling state
under the emerging time evolution, whereas it is not so in the
Bogoliubov state, which also breaks supersymmetry.

\medskip
\section{Algebraic framework}

The basic structural elements of supersymmetry are a $C^*$-algebra
$\A $  and an odd nilpotent element $Q \in \A $ (the supercharge):
\beq Q^2=0\,\Rightarrow \{Q^\dag\}^2=0 \qquad QQ^\dag + Q^\dag Q
=: H. \eeq  $H$ is supposed to be the generator of the time
evolution and by Eq.(2.1) has the properties \beq \ba{rlc}
(i) & [Q,H]=0\,\Leftrightarrow\,[Q^\dag,H]=0 & \\[4pt]
(ii) & H |0\rangle = 0 \,\Leftrightarrow \,Q|0\rangle = Q^\dag
|0\rangle = 0 & \\[4pt]
(iii) & E:\,\,\, H|e\rangle=E|e\rangle,\,\, E>0 \Rightarrow  &
\mbox{ is at least twofold degenerate.} \ea \eeq For {\it (iii)},
note that either $Q|e\rangle$ or $Q^\dag|e\rangle$ must be
different from zero and also belongs to the eigenvalue $E$. Also,
$Q|e\rangle$ cannot be $\,\sim |e\rangle$, $Q^2|e\rangle=0$ would
be in contradiction with the assumption $Q|e\rangle \not= 0$.

Equation (2.2) implies that the Hilbert space $\Ha $ can be
written as a sum of a zero-space $\Ha _0$ (projection $P_0: \Ha
_0=P_0\Ha $) and a tensor product of $\C ^2$ and the rest, $\Ha
_H$: $\Ha = \Ha _0 \bigoplus (\C^2\otimes \Ha _H )$. Defining
$\eta =Q/\sqrt H ,P_0 \eta =0 $ we have $\eta \eta^\dag +\eta^\dag
\eta =1-P_0.$ In this decomposition we can write $\eta ,H, Q $ in
a matrix representation \beq \eta =\left(\ba{ccc} 0 &  0 & 0
\\ 0 & 0 & 1 \\ 0 & 0 & 0 \ea \right),\quad H =\left(
\ba{ccc} 0 & 0 & 0 \\ 0 & H & 0 \\ 0 & 0 & H \ea \right), \quad Q
= \left( \ba{ccc} 0 & 0 & 0 \\ 0 & 0 & {\sqrt H} \\ 0 & 0 & 0 \ea
\right).\eeq Here we identify $H=H(1-P_0)$; $\sqrt H $ denotes the
positive square root of $H\in \B (\Ha _H)$, but there are others:
if $G_{\alpha }=e^{i\alpha }Q+e^{-i\alpha }Q^\dag, \alpha
\in(0,2\pi ),$ then $G_{\alpha }^2=H \, \forall \alpha$. The gauge
transformation  $G=G_0\rightarrow G _{\alpha }$ is effected by
$F=[\eta ,\eta ^\dag ]$, which has the familiar matrix
representation \beq F= \left( \ba{ccc} 0 &  0 & 0
\\ 0 & 1 & 0 \\ 0 & 0 & -1 \ea \right)\qquad G=\left(
\ba{ccc} 0 &  0 & 0 \\ 0 & 0 & \sqrt{H} \\ 0 & \sqrt{H} & 0 \ea
\right),\eeq so that $e^{i\alpha F}Ge^{-i\alpha F}=G_{\alpha}$.

In $\C^2\otimes \Ha _H $ every element can be written as $A\eta
\eta^\dag +B \eta ^\dag \eta +C\eta +D\eta^\dag , A,B,C,D \in \B
(\Ha _H)$. This gives the algebra a grading by calling the first
two terms even and the others odd. All this emerges from a single
nilpotent operator, namely $Q$. The operators $G_{\alpha }$
generate a supertransformation \beq a_k\,\ra\, a_k(s, \alpha ) =
e^{isG_{\alpha }}a_ke^{-isG_{\alpha }}, \eeq which mixes even and
odd elements of $\A$. To isolate the various aspects, we start
with a finite-dimensional $\A$, $i,k = 1,2,...,N$ and later
investigate the limit $N\ra\infty$.

\medskip
\section{Fermions on a lattice}

The supertransformation is a non-linear transformation of the
$a$'s that preserves their algebraic relations. For warming up we
start with the simplest case:
$$
N=1,\,\,Q=a,\,\,G=e^{i\alpha }a+e^{-i\alpha }a^\dag.
$$

\noindent The supertransformation is \beq a(s,\alpha ) = (\cos
s)^2a + e^{2i\alpha }(\sin s)^2a^\dag  +ie^{i\alpha } \cos s \sin
s (a^\dag a-a a^\dag ). \eeq In this special case $H$ has to be
trivial since it is twofold degenerate, {\it (ii)} cannot appear
since $a, a^\dag$ create all of $\A$. The supertransformation
$a\rightarrow a(s,\alpha )$ is a two-dimensional generating subset
(but not subgroup) of the automorphism group. The latter is
three-dimensional and isomorphic to $SU(2)$. Equation (2.3) tells
us that in its embryonic form the supertransformation is a
Bogoliubov transformation plus a quadratic term.

To get such an explicit expression also for higher $N$ we restrict
the systems to be considered first by a locality condition. We
think in terms of a lattice and assume $Q$ to be the sum of
charges situated at the lattice sites: \beq Q=\sum_{i=1}^N
q_i.\eeq Equation (2.1) requires $\{q_i,q_j^\dag \}=0 \,\,\forall
i\neq j$, in addition to $\{q_i,q_j \}=0$.

\vspace{10pt} \noindent\Un{{\bf I.} One kind of fermions at each
lattice site}

\medskip
The most general $Q$ is of the form \beq Q=\sum_i z_i a_i,\,\,
z_i\in {\bf C}. \eeq Since the third power of a fermion operator
at a point vanishes, $q_i$ must be linear in $a$ and $a^\dag$, and
the nilpotence leaves only $a$ or $a^\dag$ (which one does not
matter). The phase of the $a_i$ is arbitrary, so we may take $z_i
\in {\bf R^+}$, that is $\alpha _i =0$. Thus \beq G=\sum_i z_i(a_i
+ a_i^\dag)\,,\quad H=\sum_i z_i^2, \eeq so with $H$ being a
$c$-number the time evolution is trivial. The $\Ha _0$ introduced
in Section 2 is empty, a vector $|0\rangle$ with
$Q|0\rangle=Q^\dag |0\rangle=0$ would be annihilated by all $a_i $
and $a^\dag _i$. Since they span all of $\A $ such a vector must
be zero. The $\eta $ of Section 2 is $\sum _i z_i a_i/(\sum _k
z_k)^{-1/2}$ and is a collective fermion coordinate. Since it
obeys the CAR-relations we have $||\eta ||=1$ although it is a sum
of $N$ operators with of a norm $N^{-1/2}$.

However, the supertransformation $e^{isG}$ is not trivial; also it
is not just a tensor product of the unitaries of the baby model
since the $q_i$ and $q_i^\dag$ in (3.2) at different points
anticommute. We find rather
$$
G a_k = z_k - a_k G\,\ra\, G^na_k = \frac{1-(-1)^n}{2}z_k G^{n-1}
+ (-1)^n a_k G^n
$$
$$
\Rightarrow \, e^{isG}a_k = a_k e^{-isG} +
z_k\,\frac{e^{isG}-e^{-isG}}{2G}\,,
$$
so that \beq \ba{l} a_k(s) = e^{isG} a_k e^{-isG} = (a_k -
z_k/2G)\,e^{-2isG} + z_k/2G \ea, \eeq where we use the notation
$a_k(0)=a_k.$ One readily verifies that this is an automorphism of
$\A$,
$$
a_k^2(s) = 0\,,\quad \{a_k(s), a_j^\dag(s)\}=\delta_{kj}\,,
\quad\{a_k(s), a_j(s)\} = 0
$$
but it is not a local transformation  --- $a_k(s)$ depend on all the
other $a$'s and $a^\dag$'s.

\vspace{10pt}

\noindent\Un{{\bf II.} Two fermions at each lattice site}

\medskip
We think of it as of electrons with spin up and down. Thus $\A$ is
generated by $a_{\uar,i}$ and $a_{\dar,i}$. So now we afford at
each site a product of three operators, of which there are four
types:
$$
a_\uar^\dag a_\uar a_\dar\,,\,\, a_\uar^\dag a_\dar a_\dar^\dag\,,\,\,
a_\uar a_\dar^\dag a_\dar\,,\,\, a_\uar a_\uar^\dag a_\dar^\dag\,.
$$
Each of them is nilpotent. A typical local supercharge is \beq Q =
\sum_i a_{\uar,i}^\dag a_{\uar,i} a_{\dar,i}\,z_i\,, \,\, z_i \in
{\bf R^+} \eeq and \beq G = \sum_i z_i a_{\uar,i}^\dag a_{\uar,i}
(a_{\dar,i} + a_{\dar,i}^\dag). \eeq A priory, $H=G^2$ seems to be
of 6-th power in the $a$'s but by locality it can be only quartic
and $(a^\dag a)^2 = a^\dag a$ reduces it to something quadratic:
\beq H=\sum_i z_i^2\,a_{\uar,i}^\dag a_{\uar,i} .\eeq Thus a free
time evolution where half of the fermions are quiet is
supersymmetric with a local supercharge. In this case the vacuum
$|\uar 0\rangle, a_{\uar,i}|\uar 0\rangle$ satisfies
$$
Q|\uar 0\rangle = Q^\dag |\uar 0\rangle = 0,
$$
irrespective of the down spins. All these vectors belong to the
eigenvalue zero of $H$ and span $\Ha _0.$ In fact the eigenvalues
are at least $2^N$-fold degenerate.

We determine the automorphism of $\A$ generated by the
supertransformation $e^{isG}$ by the same method as before:
$$
Ga_{\uar,k} = z_k a_{\uar,k}(a_{\dar,k} + a_{\dar,k}^\dag) -
a_{\uar,k}G\,\ra\, G^n a_{\uar,k} = (-1)^n a_{\uar,k} (G -
z_k(a_{\dar,k} + a_{\dar,k}^\dag))^n,
$$
leading to
\beq
a_{\uar,k}(s) = a_{\uar,k} e^{-is(G-(a_{\dar,k} +
   a_{\dar,k}^\dag)z_k)} e^{-isG}.
\eeq
Similarly,
$$
G a_{\dar,k} = z_k a_{\uar,k}^\dag a_{\uar,k} - a_{\dar,k}G\,\ra\,
G^n a_{\dar,k} = (-1)^n a_{\dar,k} G^n + \frac{1-(-1)^n}{2} z_k
a_{\uar,k}^\dag a_{\uar,k} G^{n-1},
$$
leading to \beq a_{\dar,k}(s) = a_{\dar,k} e^{-2isG} + z_k
a_{\uar,k}^\dag a_{\uar,k} \frac{1-e^{-2isG}}{2G}. \eeq Of course
there is an alternative form where $e^{2isG}$ is pulled out to the
left and with which one verifies that this strange transformation
$\,a_{\uar\dar,k}\ra a_{\uar\dar,k}(s)\,$ is actually an
automorphism group of $\A$ that mixes spin up and down as well as
even and odd; however, the time evolution leaves $a_{\dar ,k}$
invariant up to a phase and does not mix between even and odd
elements or between elements at different sites.

\medskip \noindent{\bf Remark}

\noindent Instead of an opposite spin one might use the next
neighbour and try \beq Q = \sum_i a_i^\dag a_i a_{i+1} z_i, \eeq
but this is not nilpotent. To meet this condition more refined
constructions are necessary (see, e.g. \cite{F}).

\vspace{10pt}

\noindent\Un{{\bf III.} Three fermions at each lattice site}

\medskip

In the case of three fermions at each lattice site, we start again
with the CAR-algebra $\A$ generated by $\{a_i^\alpha\}$: \beq
\{a_i^\alpha, {a^\dag_k}^\beta\} = \delta^{\alpha
\beta}\delta_{ik},\,\,\{a_i^\alpha, a_k^\beta\} = 0, \qquad i,k =
1,\dots,N,\,\, \alpha, \beta = 1,2,3. \eeq However even with
strictly local $q_i$ the charge $Q=\sum _i^N q_i$ in (3.2) creates
a non-local supertransformation, so we drop locality. Instead we
impose translation invariance of the $q_i$'s in such a way that
$Q$ becomes translation- and even permutation- invariant.
Furthermore we think of the $a_i^1$ and $a_i^2$ as Cooper pairs,
and thus consider the subalgebra $\C$ of $\A$ generated by $b_i =
a_i^1 a_i^2$ and $a_k^3$, $i,k = 1,\dots, N$. Although the $b_i$'s
commute for different sites they do not form a {\sl bona fide}
Bose field since there is at most one pair per site, $b^2=0$.
However in $\C$ the anticommutator $\{b_i^\dag, b_i\}$ is a
projection of the centre and therefore in an irreducible
representation it equals unity. These are the representations we
are interested in and therefore we can think of the $b$'s as of
spin variables: $$ b_i = \frac{\sigma_i^x-i\sigma_i^y}{2},\qquad
1-2b_i^\dag b_i = \sigma_i^z.$$ Thus our algebra $\C$ is defined
by \beq \ba{l}\{a^3 _i, a^{3\dag} _j\}=\delta_{ij},\,\{a^3 _i,
a^3 _j\}=0,\,[a^3 _i, b_k]=0\\[6pt] [b_i,
b_k^\dag]=\delta_{ik}(1-2b_i^\dag b_i),\quad \{b_i, b_i^\dag\}=1.
\ea\eeq

The supertransformation, and therefore the dynamics, will be
defined by fluctuation variables.

\vfill\eject

\noindent {\bf Definition}
\medskip
\beq M_N = \frac{1}{\sqrt N}\sum_{k=1}^N b_k,\qquad
\eta_N=\frac{1}{\sqrt N}\sum_{k=1}^N a^3_k. \eeq

\bigskip
\noindent {\bf Proposition}

\medskip
\noindent
\begin{itemize}
\item[{\it (i)}] $\eta_N\eta_N^\dag + \eta_N^\dag \eta_N = 1,
\quad \eta_N^2 = 0$; \item[{\it (ii)}] $[M_N, M_N^\dag] =
1-\dfrac{2}{N}\sum_{k=1}^N b_k^\dag b_k$; \item[{\it (iii)}]
$[M_N, \eta_N]=[M_N, \eta_N^\dag]=0$.
\end{itemize}

\bigskip
\noindent {\bf Remarks}

\begin{enumerate}
\item $\eta$ represents a collective Fermi mode and will serve as
a Clifford variable. The fact that the number of single fermion
modes $s$ equals the number of pairs is not essential; \item $M$
is a collective Bose mode and, in a representation based on the
``vacuum" $|0\rangle : \quad b_i|0\rangle=0$ (all spins down), it
assumes for $N\ra\infty$ the properties of $(x+ip)/\sqrt 2$ in
quantum mechanics and we will arrive at Witten's supersymmetric
quantum mechanics \cite{Wit}; \item By anticommutativity the
$a^3_k$ are so correlated that $\Vert \eta_N \Vert =1 \,\,\forall
N$. On the contrary, $\Vert M_N \Vert =\sqrt N /2$ since we can
think of $M_N$ as of $(S^x-i S^y)/(2\sqrt N)$, with
$S=\sum_{k=1}^N \sigma_k$, $b=(\sigma^x-i\sigma^y)/2$.
\end{enumerate}

In agreement with our desideratum $Q=\sum_i q_i$, $\{q_i,
q_j\}=0$, we take $q_i=b_i\eta/\sqrt N$ and thus obtain \beq
\ba{l}Q_N =
M_N\eta_N\\[6pt]
G=Q_N+Q_N^\dag\\[6pt]
H_{SS}=G^2=\{Q_N,Q_N^\dag\}=\\[6pt]
\qquad =\eta_N\eta_N^\dag M_N M_N^\dag + \eta_N^\dag\eta_N
M_N^\dag M_N = M_N^\dag M_N + \eta_N\eta_N^\dag
\left(1-\dfrac{2}{N}\sum_{k=1}^N b_k^\dag b_k\right). \ea\eeq

\bigskip
\noindent {\bf Remarks}

\begin{enumerate}
\item In the BCS model (in the degenerated case) $H_{\rm BCS} =
-M_N^\dag M_N\,$ and we see that it differs from $-H_{SS}$ only by
$O(1)$. Thus the energies per particle $H/N$ coincide for
$N\ra\infty$; \item In the $S_z=0$ situation, where $M_N \st
{N\ra\infty}{\ra} (x+ip)/\sqrt 2$ with Pauli matrices for
$\eta=(\tau_x+i\tau_y)/2$, we have $H_{SS}=(x^2+p^2)/2+\tau_z/2$.
\end{enumerate}

Next we shall briefly comment on the ground state and the ceiling
state of $H_{SS}.$ For their discussion it has to be kept in mind
that, for $N\ra\infty$, $\eta_N$ stays bounded and only $M_N$ can
grow big; $H_{SS}$ is then essentially
$(S^x)^2+(S^y)^2=S^2-(S^z)^2$. The smallest $H_{\rm SS}$ requires
$S^z$ as big as $S$. But the lowest $H_{\rm BCS}=-H_{\rm SS}$
wants $S^z$ close to zero and maximal $S$. The Fock vacuum
$b|0\rangle=0$ gives $H_{\rm SS}=0$ if the single fermions are
anticorrelated to the pairs, $\eta^\dag|0\rangle=0$. Then
$G|0\rangle = (M_N\eta_N + M_N^\dag \eta_N^\dag)|0\rangle =0$ and
thus $Q_N|0\rangle = Q_N^\dag|0\rangle = H_{\rm SS}|0\rangle =0$.
For even $N$ there are more ground states if all pairs are
pair-wise anticorrelated. By this we mean that there is a
permutation $p_i$, $i=1,\dots,N/2$ such that $b_i|0\rangle =
-b_{i+p_i}|0\rangle$ and $b_i^\dag |0\rangle =
-b_{i+p_i}|0\rangle$. Then $M_N|0\rangle = M_N^\dag |0\rangle =0$
and again $Q_N|0\rangle = Q_N^\dag |0\rangle =0$, however $\eta_N$
acts on $|0\rangle$. In this case $S=S^z=0.$

For the ground state of $-H_{\rm SS}$ we want $S^z=0$, $S=N$. For
even $N$ this is possible by applying $N/2$ times $M_N^\dag$ onto
$|0\rangle$. For big $N$ this becomes awkward and here it is
expedient to make a Bogoliubov transformation. However the
standard form $$ a^1 \ra \frac{a^1+a^{2\dag }}{\sqrt 2}, \qquad
a^2 \ra \frac {a^2-a^{1 \dag }}{\sqrt 2}, $$ does not leave the
pair algebra $\C\subset\A$ invariant and we have to use the
transformation (2.1) with $\cos s =0$, $\sin s =1$, i.e. $$ b \ra
\frac{b-b^\dag}{2}+\frac{1}{2}-b^\dag b,$$ such that $M_N$ becomes
$$ M_N = \frac{\sqrt N}{2}+\frac{1}{\sqrt N}\sum_i\left[b_i^\dag b_i
+\frac{1}{2}(b_i-b_i^{\dag })\right]. $$ If $|0\rangle $ denotes
the new $b$-vacuum we note $||M_N-\frac {\sqrt
N}{2}|0\rangle||^2=1/4.$ We conclude \beq \lim _{N\rightarrow
\infty }||\frac{4}{N}M^{\dag }_N M_N||=1.\eeq

\medskip
\section{The limit $N\ra\infty$}

In the limit $N\rightarrow \infty$, new features appear. We have
to distinguish between local, mesoscopic and macroscopic
observables: typically $\vec{\sigma }^j, \lim _{N\rightarrow
\infty }\frac { \vec{S}}{\sqrt N}, \lim _{N\rightarrow \infty
}\frac{ \vec{S}}{N}.$ Which limits exist and how they behave under
the time evolution (TE) and the supertransformation (ST) will
depend very much on the state.

For the limiting procedure we impose only minimal requirements. We
assume that a state for arbitrary $N$ is given and we check
whether the expectation values converge. In addition we demand
that these limits can be interpreted as the expectation values of
a limiting algebra. We want the latter to be as big as necessary
for the mesoscopic observables to still reflect some quantum
features.

Let us first turn to the global unscaled quantities \beq \ba{lcl}
S_\alpha = \dsum_{k=1}^N \sigma_\alpha^k & \quad & S_\pm =
\dfrac{1}{2}(S_x\pm iS_y)\\[6pt] [S_+, S_-] = S_z & &
[S_z, S_\pm] = \pm 2S_\pm. \ea\eeq In terms of these operators
together with $\eta $ from (3.14) in the model of Section 3:
\beq\ba{l} G_\alpha =\dfrac{1}{\sqrt N}
(e^{i\alpha}\eta S_- + e^{-i\alpha}\eta^\dag S_+)\\[6pt]
H_{\rm SS} = G_\alpha^2 = \dfrac{1}{N}\{S_+S_- + [S_-,\, S_+]\eta
\eta^\dag\} \ea\eeq

$H_{\rm SS}$ is independent of the gauge transformation
$\gamma_\alpha G = G_\alpha$. For the time evolution (with $\dot A
= -i[A, H]\,$) this leads to \beq\ba{l} \dot\sigma_z^{(j)} =
-i\sigma_+^{(j)}\dfrac{S_-}{N} +
i\dfrac{S_+}{N}\sigma_-^{(j)}\\[4pt] \dot\sigma_+^{(j)} = -i
\dfrac{S_+}{N}\sigma_z^{(j)}-i\frac{2\sigma ^{(j)}_+ \eta
\eta^{\dag}}{N}
\\[4pt] \dfrac{\dot S_+}{N} = -i
\dfrac{S_+S_z}{N^2}- \frac{2S_+}{N^2}\eta \eta^{\dag}\quad\quad
\dot\eta = -i\dfrac{\eta}{N}[S_-, S_+]\quad\quad {\dot
S_z}=0,\ea\eeq whereas for the supertransformation ($A' = -i[A,
G_\alpha]\,$) we obtain \beq\ba{l} \sigma_z^{(j)'} =
2ie^{i\alpha}\eta\dfrac{\sigma_-}{\sqrt N}
-2ie^{-i\alpha}\eta^\dag\dfrac{\sigma_+}{\sqrt N}\\[4pt]
\sigma_+^{(j)'} = -ie^{i\alpha}\eta\dfrac{\sigma_j^{(j)}}{\sqrt
N}\\[4pt] \dfrac{S_+'}{N} = ie^{i\alpha}\eta\dfrac{S_z}{N\sqrt
N}\quad\quad \eta' = i\dfrac{S_+}{\sqrt N}[\eta, \eta^\dag]
\\[4pt] \dfrac{S_z'}{N}=2i\frac{1}{N\sqrt N }(e^{i\alpha}
\eta S_- -e^{-i\alpha }\eta ^{\dag} S_+).\ea\eeq

For $N\ra\infty$ evidently $\lim \vect{\sigma}^{(j)'} = 0$. For
the time evolution we can use the fact that $\vect{S}/N$ is a norm
bounded sequence. Therefore if $N\ra\infty$ the time derivatives
will have weak accumulation points. To be able to construct a
corresponding automorphism group, however, we need strong
convergence that will only hold in favourable representations.
Especially $\vect{S}/N$ will be in the centre of the
representation, and supersymmetry becomes trivial in local and
global operators $\vect{\sigma }$ and $\vect{S}/N$.

\vfill\eject {\bf The ground state of $H_{\rm SS}$}

\medskip
 The ground state is given as the expectation value with the
``vacuum vector" $|0\rangle $ with all spins down:
$\,S_z^{(N)}|0\rangle = -N|0\rangle$,
${(\vect{S}^{(N)})^2}|0\rangle = N(N+2)|0\rangle$. This means for
quasi-local operators: \beq \omega (\Pi _j \sigma ^j_{k_j})=\Pi
_j(-\delta _{k_j,3}). \eeq The expectation value factorizes and is
the same for all $k$. Following \cite{GVV} we can extend the set
of observables and introduce the fluctuation operators\beq
W_N(\alpha,\beta):=\exp\left\{i\sum_{k=1}^N \frac{\alpha\sigma_x^k
- \beta\sigma_y^k}{\sqrt 2N}\right\}.\eeq Hence
$$\ba{ccl} \langle 0|W_N(\alpha, \beta)|0\rangle & = & \langle
0|\dar\dots\dar|1-\frac{\alpha^2+\beta^2}{4N} +
O(N^{-2})|\dar\dots\dar\rangle^N\\[12pt] & \longrightarrow &
e^{-(\alpha^2+\beta^2)/4} = \langle 0|e^{i(\alpha q+\beta
p)}|0\rangle,\ea$$ where the vector $|0\rangle$ is now the ground
state of a harmonic oscillator over a Weyl algebra $\W$ generated
canonically by $x$ and $p$. If in addition to $W_N(\alpha, \beta)$
we have a local polynomial $\prod_{j=1}^M \sigma_{\alpha_j}^j$,
then for fixed $M$ and $N\ra\infty$ the expectation value
factorizes since the interference between the two factors vanishes
as $1/\sqrt N$.

Furthermore
$$\ba{ccl}\langle 0|W_N(\alpha,0) W_N(0,\beta)|0\rangle & = & (\langle
\dar\dots\dar |(1+i\frac{\alpha\sigma_x}{\sqrt 2N} -
\frac{\alpha^2}{4N})(1 - i\frac{\beta\sigma_x}{\sqrt 2N} -
\frac{\beta^2}{4N})|\dar\dots\dar\rangle)^N\\[12pt]
& \ra &(1 - \frac{\alpha^2+\beta^2}{4N} +
i\frac{\alpha\beta}{2N})^N \ra \langle 0|W(\alpha, \beta)|0\rangle
e^{-i4\alpha\beta},\ea$$ so the Weyl relations hold in the limit
$N\ra\infty$, which allows us to call $\lim _{N\rightarrow
\infty}W_N(\alpha, \beta )=W(\alpha, \beta )=e^{i\alpha q+\beta
p}.$

Thus $N\ra\infty$ maps our operators into the factorizable, hence
commuting product $\W\otimes\A_{\rm loc}$ such that $S_z/N\ra -1$,
$S_\pm/\sqrt N \ra(q\mp ip)/\sqrt 2$. The evolution equations
(4.3),(4.4) become (with $\eta_\alpha = e^{i\alpha}\eta$): \beq
\dot\sigma_{\alpha_k}^k\ra 0,\qquad \dot q \ra p,\qquad \dot p \ra
-q,\qquad \dot\eta_\alpha\ra i\eta_\alpha, \eeq \beq
(\sigma_{\alpha_k}^k)'\ra 0,\quad
q'=\dfrac{\eta_\alpha-\eta_\alpha^\dag}{\sqrt 2},\quad
p'=\dfrac{\eta_\alpha+\eta_\alpha^\dag}{\sqrt 2},\quad
\eta_\alpha'=\dfrac{q-ip}{\sqrt2}\,[\eta_\alpha ^{\dag},\,
\eta_\alpha].\eeq They are indeed implied by the limiting
generators \beqa H =
(q^2+p^2-1)/2 + \eta_\alpha\eta_\alpha^\dag,\\[6pt]G_\alpha =
\eta_\alpha(q+ip)/\sqrt 2
 + \eta_\alpha^\dag (q-ip)/\sqrt 2.\eeqa
Notice, however, that the evolution of the global operators is not
the limit of the (norm) evolution of the local ones \cite{HN}. The
local $\sigma$'s remain constant, but the mesoscopic $q$ and $p$
move.

\bigskip \noindent{\bf Remark}

\noindent In the matrix representation
$$\eta_\alpha=\left(\ba{ll}0 & e^{i\alpha}\\[2pt]0 & 0\ea\right),$$
the ground state of $G_{\alpha}$, $\eta_\alpha^\dag|0\rangle
=(q+ip)|0\rangle = 0$, is given by the vector
$$\left|\ba{l}0\\[2pt]e^{-x^2/2}\ea\right\rangle.$$
It is thus the same for all $\alpha$, so there is no symmetry
breaking. From the well known relation for the eigenvectors of a
total spin $\vec{S}^2$ and $S_z$, expanding around $s_z=-s$, \beq
S_z|s,s_z\rangle =s_z|s,s_z\rangle, \quad S_{\pm }|s,s_z\rangle
=\sqrt {s(s+1)-s_z(s_z\pm 1)}|s,s_z\pm 1\rangle, \eeq it follows
that for all spins in the $z$-direction the macroscopic $S_z/N$
and the mesoscopic $S_x/\sqrt N,S_y/\sqrt N$ converge.

\vspace{0.7cm} {\bf The ceiling state of $H_{\rm SS}$}

\medskip
Next we want to examine the ground state of $H_{\rm BCS} = -H_{\rm
SS}$, which is essentially $S_z^2-S^2$. Thus $S$ should be as big
as possible and $S_z=0$. Since for each eigenvalue $E$ of $H$
there is an eigenvector to $G_\alpha$ with eigenvalue $\sqrt E$,
we have to find a maximal eigenvector $|\Omega_\alpha\rangle$ to
$G_\alpha$. This vector will depend on $\alpha$ but all
$|\Omega_\alpha\rangle$ are equally suitable for $H_{\rm BCS}$.
Therefore the ground state of $H_{\rm BCS}$ is degenerate.

To get the limiting state we have to find a sequence
$|\Omega_{\alpha,N} \rangle$ such that \beq \Vert(G_\alpha -
E_N)\Omega_{\alpha, N}\Vert = 0.\eeq

With the notation
$$
\eta =\left( \ba{cc} 0 & 1 \\ 0 & 0 \ea \right),
$$
we can write the operators $G_\alpha$ and $H_{\rm SS}$ in matrix
form as follows: \beq\ba{l}G_\alpha = \left(\matrix{0 & e^{i\alpha}S_+\\[2pt]
e^{-i\alpha}S_- & 0}\right)\,,\\[12pt]
H_{\rm SS} = \left(\matrix{S_+S_- & 0\\[2pt] 0 & S_-S_+}\right) =
\frac{1}{4}\left(\matrix{S^2-S_z^2+2S_z & 0\\[2pt]
0 & S^2-S_z^2-2S_z}\right)\,.\ea\eeq

An eigenvector has to satisfy
\beq\ba{rcl}\left(\matrix{0 & e^{i\alpha}S_+ \\[2pt] e^{-i\alpha}S_- & 0}\right)
\left|\matrix{\psi_1 \\[2pt] \psi_2}\right\rangle & = &
\left|\matrix{e^{i\alpha}S_+\,\psi_2\\[2pt]e^{-i\alpha}S_-\,\psi_1}\right\rangle
= \sqrt E\,\left|\matrix{\psi_1\\[2pt]\psi_2}\right\rangle\\[16pt]
S_-|\psi_{1,\alpha}\rangle &=& e^{i\alpha}\sqrt E |\psi_{2,\alpha}\rangle\\[4pt]
S_+|\psi_{2,\alpha}\rangle &=& e^{-i\alpha}\sqrt E
|\psi_{1,\alpha}\rangle.\ea\eeq

Evidently, we can choose $|\psi_{1,\alpha}\rangle=|\psi_1\rangle$,
$|\psi_{2,\alpha}\rangle=e^{-i\alpha}|\psi_2 \rangle $ and all
$\alpha$ lead to
$$\ba{lcl}
(S^2-S_z^2+2S_z)|\psi_{1,N}\rangle &=&4E_N|\psi_{1,N}\rangle\\[8pt]
(S^2-S_z^2-2S_z)|\psi_{2,N}\rangle&=&4E_N|\psi_{2,N}\rangle .
\ea$$

With an appropriate symmetrization we need
$$\ba{l}|\psi_{1,N}\rangle=\frac{1}{\sqrt 2}\,\vert
(\uar)^{\frac{N}{2}+1}(\dar)^{\frac{N}{2}-1}\rangle\\[8pt]
|\psi_{2,N}\rangle=\frac{1}{\sqrt 2}\,\vert
(\uar)^{\frac{N}{2}}(\dar)^{\frac{N}{2}}\rangle\,.\ea$$ Then
$$\ba{l}(S_z^2-2S_z)|\psi_{1,N}\rangle=(4-4)|\psi_{1,N}\rangle=0\\[8pt]
(S_z^2+2S_z)|\psi_{2,N}\rangle=0.\ea$$

Thus for finite $N$ we have
$$
|\psi_1\rangle=S_+|\psi_2\rangle,\quad |\psi_2\rangle=S_+^{N/2}|0\rangle,
\quad 4E_N=N(N+2).
$$
This form of $|\psi\rangle$ does not lend itself easily to
$N\ra\infty$, but we can write $|\psi_2\rangle$ as an integral
over vectors with all spins pointing in the $(\cos\alpha,
\sin\alpha,0)$-direction.

\bigskip
\noindent{\bf Proposition} \beq\ba{l}|\psi_2\rangle=\C
\lint_{-\pi/2}^{\pi/2}d\alpha\,e^{i\alpha
S_z}\,e^{i(\pi/2)S_y}|0\rangle=:\C\lint_{-\pi/2}^{\pi/2}d\alpha\,
|\alpha\rangle\\[12pt]
\C=\left(\pi\,\lint_{-\pi/2}^{\pi/2}d\vp\cos^N\vp\right)^{-1/2}\,.\ea\eeq

\bigskip
\noindent{\bf Proof}

\noindent The integrand is periodic with period $\pi$; thus,
changing $e^{i\alpha S_z}$ to $e^{i(\alpha+\vp)S_z}$ does not
alter the integral. Hence, $e^{i\vp
S_z}|\psi_2\rangle=|\psi_2\rangle$ and $S_z|\psi_2\rangle=0$. Now
$$\ba{ccl}\langle\psi_2|\psi_2\rangle &=&
\C^2\lint_{-\pi/2}^{\pi/2}\langle \ra\dots\ra|e^{-i\vp'
S_z}\,e^{i\vp S_z}|\ra\dots\ra\rangle d\vp d\vp'\\[6pt]
&=& \C^2\lint_{-\pi/2}^{\pi/2}d\vp
d\vp'\cos^N(\vp-\vp')\,=\,\C^2\pi\lint_{-\pi/2}^{\pi/2}d\vp\cos^N\vp\,=\,1.\ea$$

\bigskip
Coherent superpositions of eigenvectors of the supersymmetry
operator $G_\alpha$ lead to the same state in the quasi-local
algebra. This state is an integral over product states. The
corresponding von Neumann algebra in the GNS representation has a
non-trivial centre. The elements of the centre correspond to the
orientation in the $x$--$y$ plane of the pure product states.

Note that although we started with a coherent superposition the
limiting state over the local algebra appears to be a mixed state:
$$\ba{ccl}C^2\lint_{-\pi/2}^{\pi/2}d\alpha
d\alpha'\langle\alpha_M|\dprod_{k=1}^M \sigma^k|\alpha'\rangle &=&
\lint_{-\pi/2}^{\pi/2}d\alpha
d\alpha'\dfrac{\cos^{N-M}\alpha'}{\lint_{-\pi/2}^{\pi/2}d\vp
d\vp'\cos^N(\vp-\vp')}\,\langle\alpha_M|\prod_{k=1}^M\sigma^k|\alpha'_M\rangle\\[12pt]
&\ra&\dfrac{1}{\pi}
\lint_{-\pi/2}^{\pi/2}d\alpha\,\langle\alpha_M|\prod_{k=1}^M\sigma^k|\alpha_M\rangle\,.\ea$$

However, we are not interested only in the state over the
quasi-local algebra. For global operators we get good convergence
properties:
$$\ba{l} \left\Vert\left(\dfrac{1}{N}\right)^k
\left(\matrix{(S_+)^k & 0\\[4pt] 0 &
(S_+)^k}\right)\left\vert\matrix{\psi_{1,N}\\[4pt]\psi_{2,N}}\right\rangle -
\dfrac{1}{2\sqrt2}\left\vert\matrix{(\uar)^{\frac{N}{2}+k}(\dar)^{\frac{N}{2}-k}\\[4pt]
(\uar)^{\frac{N}{2}+k+1}(\dar)^{\frac{N}{2}-k-1}}\right\rangle\right\Vert=0\\[18pt]
\lim\dfrac{1}{2}\left\langle\matrix{(\uar)^{\frac{N}{2}+k}(\dar)^{\frac{N}{2}-k}\\[4pt]
0}\right\vert\left(\matrix{S_z & 0\\[4pt]0 & S_z}\right)
\left\vert \matrix{(\uar)^{\frac{N}{2}+k}(\dar)^{\frac{N}{2}-k}\\[4pt]0}\right\rangle = 2k.\ea$$

We can therefore interpret the limits of the expectation values as
expectation values of operators acting in the Hilbert space
$L^2[T,d\vp)\otimes C^2$, with \beq\ba{cclcccl}\eta &
\longrightarrow & 1\otimes\left(\matrix{0 & 1\\[4pt] 0 & 0}\right)
& \qquad\qquad & \dfrac{S_+}{N} & \longrightarrow & e^{2\pi
i\vp}\\[12pt]
S_z & \longrightarrow & 2p_\vp & & \dfrac{S_-}{ N} &
\longrightarrow & e^{-2\pi i\vp},\ea\eeq where $p_\vp$ is the
momentum on $T$ with periodic boundary conditions.

\medskip
\section{Comparison with the BCS theory}

Without referring to supersymmetry, we consider the Hamiltonian
\beq H'_{\rm BCS} = -\frac{S_+S_-}{N} = -\frac{S^2-S_z^2}{4N} =
-\frac{S_x^2+S_y^2}{4N}\,.\eeq This operator differs from the
previous one only by $S_z/N$, which is a bounded operator and
therefore converges to an operator in the centre, which does not
affect the dynamics. Thus the two Hamiltonians can be thought of
as describing one and the same physical situation.

In this case the ground-state energy can be approximated by the
rotated ground state of $H_{\rm SS}$, namely by \beq
|\psi_{\alpha,N}\rangle = \prod\left(\frac{1}{\sqrt2}\right)^N
\left| \ba{c} e^{i\alpha } \\ e^{-i\alpha }\ea \right\rangle .\eeq

For the limits of the expectations of the local operators we
obtain \beq\ba{l}\lim\langle\psi_{\alpha,N}|\prod
\sigma_{\alpha_j}^{i_j}|\psi_{\alpha,N}\rangle = \dprod_j
\left(\frac{1}{2}\right)^N
\left\langle\matrix{e^{i\alpha}\\[2pt]e^{-i\alpha}}\right|
\sigma_{\alpha_j}^{ij}\left|\matrix{e^{i\alpha}\\[2pt]e^{-i\alpha}}
\right\rangle,\ea\eeq which is a pure product state.

To evaluate the global operators in this scaling limit, we must
adjust the renormalization to the parameter $\alpha$. In
particular, for $\alpha=0$:
$$
\ba{l}\lim\dsum_k\dfrac{\sigma_x^k}{N}=1\\[8pt]
\lim\langle\psi_{0,N}|A_{\rm
loc}\left(\dfrac{\sum(\sigma_x^k-1)}{\sqrt
N}\right)^2 A_{\rm loc}|\psi_{0,N}\rangle = 0\\[8pt]
\lim\langle\psi_{0,N}|A_{\rm loc}\,e^{ir\sum(\sigma_x^k-1)/\sqrt
N}\,A_{\rm loc}|\psi_{0,N}\rangle = 1\,.\ea
$$

For $\sigma_y$ and $\sigma_z$ we get
$$\ba{l}
\lim\langle\psi_{0,N}|\dfrac{\sum \sigma_y^k}{\sqrt N}
|\psi_{0,N}\rangle = 0\\[8pt]
\lim\langle\psi_{0,N}|\dfrac{(\sum \sigma_y^k)^2}{N}
|\psi_{0,N}\rangle = 1\\[8pt]
\lim\langle\psi_{0,N}|e^{ir\sum\sigma_y^k/\sqrt N}
|\psi_{0,N}\rangle = e^{-\frac{r^2}{2}}\\[8pt]
\lim\langle\psi_{0,N}|e^{ir\sum\sigma_z^k/\sqrt N}
|\psi_{0,N}\rangle = e^{-\frac{r^2}{2}}\,. \ea$$

Similarly to the ground state of $H_{\rm SS}$, we can interpret
\beq\ba{l} \lim e^{ir\sum\sigma_y^k/\sqrt N} = e^{irq}\\[8pt]
\lim e^{is\sum\sigma_z^k/\sqrt N} = e^{isp}\\[8pt]
\lim e^{ir\sum\sigma_y^k/\sqrt N}e^{is\sum\sigma_x^k/\sqrt N}
e^{-ir\sum\sigma_y^k/\sqrt N} = e^{irs}e^{isp}\,, \ea\eeq as a
non-trivial global algebra, so that $e^{irq}$ and $e^{isp}$
satisfy the Weyl relations in $L^2(R, dq)$, and $q$ and $p$ can be
interpreted as space and momentum operators respectively. This is
nothing else but the fluctuation algebra as discussed in
\cite{GVV}.

The time evolution however shows new features. On the quasi-local
algebra it corresponds to a rotation of the individual spins
around the $x$-axis \cite{TW}. For the global algebra we get
something quite different. In leading order in $N$ we have \beq
\left[ \frac{S_+S_-}{N},\frac{S_x-N}{\sqrt
N}\right]=-i\frac{S_yS_z}{2N\sqrt N},\quad \left[
\frac{S_+S_-}{N},\frac{S_y}{\sqrt N}\right]=i\frac{S_xS_z}{2N\sqrt
N },\quad \left[ \frac{S_+S_-}{N},\frac{S_z}{\sqrt N}\right]=0
\eeq This corresponds, for the Weyl algebra, to
$$[H, q]= -ip, \qquad \quad [H,p]=0,$$
and thus to a free evolution, $q\ra q+pt$, $p= {\rm const}$.
Therefore no invariant state exists and the state over the
fluctuation algebra has to change in time.

Of course also in this setting we can construct a coherent
superposition of the eigenvectors $$|\psi_{g,N}\rangle =
\left\vert\int
d\alpha\,g_{\alpha,N}\psi_{\alpha,N}\right\rangle\,.$$

With the appropriate care in the normalization, we obtain on the
local level
$$\lim\left\langle\int d\alpha\,g_{\alpha,N}\psi_{\alpha,N}|A_{\rm loc}|\int
d\beta\,g_{\beta,N}\psi_{\beta,N}\right\rangle =
\int|g_\alpha|^2d\alpha\langle\psi_{\alpha,N}|A_{\rm
loc}|\psi_{\alpha,N}\rangle$$ because different $\psi_{\alpha,N}$
lead to inequivalent representations, i.e.
$$\lim\langle\psi_{\alpha,N}|\psi_{\beta,N}\rangle = 0, \qquad
\alpha\not=\beta\,.$$

The coherent superposition escapes our observation on the local
level, where it gives the same expectation value as the incoherent
superposition. If we move to the mesoscopic level, then the
fluctuation algebra has no transparent interpretation for the
statistical superposition of the states. For the coherent
superposition, especially for $g=1$, we recover the ceiling state
\beq
|\psi_{1,N}\rangle=\lint_{-\pi}^{\pi}d\alpha\,|\psi_{\alpha,N}\rangle,\eeq
and the global algebra changes from $L^2(R, dq)$ to $L^2(T, dq)$,
where the global operators are obtained in a different
scaling.\footnote{Note that a different scaling is necessary for
the weak decay properties \cite{VZ}, but here we have a
non-trivial centre and thus extend the Abelian algebra of the
centre to an irreducible algebra over the centre in which all
quantum effects are contained.}

\bigskip
\section{Summary}

We have studied three sets of operators --- local, mesoscopic and
macroscopic --- in representations based on different states of
$H_{\rm SS}$: the ground state (GS), the Bogoliubov state in the
$x$-direction (BS), and the ceiling state (CS). Recall the
limiting algebras in these three cases: \beq\ba{rl} \mbox{GS} &
\quad \left\{ \ba{l} \left(\dfrac{S_x}{2\sqrt N},
\dfrac{S_y}{2\sqrt N},
\dfrac{S_z+N}{\sqrt N}\right)\,\longrightarrow \,(q, -p, 0)\\[6pt]
\left(\dfrac{S_x}{N}, \dfrac{S_y}{N},
\dfrac{S_z}{N}\right)\,\longrightarrow \,(0, 0, -1)\ea\right.\\[18pt]
\mbox{BS} & \quad \left\{ \ba{l} \left(\dfrac{S_x-N}{\sqrt
  N}, \dfrac{S_y}{2\sqrt N}, \dfrac{S_z}{2\sqrt
  N}\right)\,\longrightarrow \, (0, -q, p)\\[6pt]
\left(\dfrac{S_x}{N}, \dfrac{S_y}{N},
\dfrac{S_z}{N}\right)\,\longrightarrow \,(1, 0, 0)\ea\right.\\[18pt]
\mbox{CS} & \quad \left\{
\ba{l} \mbox{mesoscopic operators do not converge}\\[4pt]
\mbox{macroscopic }\, \dfrac{S_\pm}{N}\,\ra\, e^{\pm i\vp}, \quad
S_z\,\ra\,2p_\vp. \ea \right. \ea \eeq In Tables 1 and 2 our
confusing results for the time evolution and for the
supertransformation respectively, for this variety of operators
are collected.

\begin{table}[htb]
$$
\ba{|l|l|l|l|} \hline & \mbox{\qquad Local} &
\mbox{\,\,\,Mesoscopic} & \mbox{\,Macroscopic} \\
\hline {\rm GS} & \vect{\sigma}^{(j)}(t)=\vect{\sigma}^{(j)} &
\ba{l}q(t)+ip(t)=e^{it}(q+ip)\\[4pt]\eta(t)=e^{it}\eta\ea
& \mbox{constant} \\ \hline {\rm BS} & \ba{l}\sigma_x^j(t)=\sigma_x^j(0)\\[4pt]
\sigma_y^j(t)+i\sigma_z^j(t)=e^{it}(\sigma_y^j+i\sigma_z^j)\ea &
\ba{l}q(t)=q-pt,\quad p(t)=p\\[4pt]\eta(t)=\eta\ea & \mbox{constant} \\
\hline {\rm CS} & \vect\sigma \mbox{ rotates around }\, (\cos\vp,
\sin\vp,
0)& \raisebox{0pt}[15pt][8pt]{$\,\,p_\vp, \vp$ constant} & \mbox{constant} \\
\hline \ea$$ \vspace{-0.5cm} \caption{The time evolution}
\vspace{0.7cm}
\end{table}

\begin{table}[hbt]
$$
\ba{|l|l|l|l|} \hline
 & \mbox{\quad Local} & \mbox{\,\,\,Mesoscopic} & \mbox{\,\,Macroscopic} \\
\hline {\rm GS} & \vect{\sigma}^{(j)}(s)=\vect{\sigma}^{(j)}(0) &
\ba{l}q'+ip'=i\eta^{\dag }\\[4pt]\eta'=i(q-ip)[\eta, \eta^\dag]\ea &
\vect{S}(s)= \mbox{constant} \\
\hline {\rm BS} & \vect{\sigma}^{(j)}(s)=\vect{\sigma}^{(j)}(0) &
\raisebox{0pt}[15pt][8pt]{$\eta'\sim \sqrt N$ becomes infinite}&
\vect{S}(s)= \mbox{constant} \\ \hline {\rm CS} & \,\sigma'\mbox{
is ill-defined} &
& \raisebox{0pt}[15pt][8pt]{$S_z'$ becomes infinite}\\
\hline \ea $$\vspace{-0.5cm} \caption{The supertransformation}
\end{table}

We have seen that the ceiling state $\omega _c$ of $H_{\rm SS}$
and the Bogoliubov state $\Omega _{\alpha }=\langle \Psi _{\alpha
}|\,.\,|\Psi _{\alpha }\rangle $ lead to the same energy per
particle. In the BCS theory, there is some discussion \cite{PM},
which one is better. We collect a few arguements to compare the
ensuing representations $\pi _c$ and $\pi _{\alpha}$.

(i) $\omega _c$ satisfies ODLRO (off-diagonal long-range order),
$\omega _{\alpha }$ does not: for $k\not= j$ \beq |\omega
_c(\sigma ^k_x\sigma ^j_x)-\omega _c(\sigma^k_x)\omega
_c(\sigma^j_x)|=1/2\eeq \beq |\omega _{\alpha }(\sigma ^k_x\sigma
^j_x)-\omega _{\alpha }(\sigma^k_x)\omega _c{\alpha
}(\sigma^j_x)|=0;\eeq

(ii) In $\pi _c$ the time evolution mixes local and mesoscopic
quantities, in $\pi_{\alpha}$ it is strictly local and corresponds
to a rotation around the $\alpha$-axis;

(iii) In $\pi _{\alpha }$ the Josephson phase is fixed to be
$\alpha$, in $\pi _c$ it is a dynamical variable;

(iv) $\pi _{\alpha }$ represents the quasi-local variables
irreducibly, in $\pi _c$ the weak closure contains the non-trivial
commuting macroscopic observables.

\bigskip
\noindent{\bf Remarks:}
\begin{enumerate}
\item ODLRO is the basis of the Meissner effect \cite{S}. The
spectrum of $p_{\vp}$ corresponds to the quantization of the
magnetic flux; \item The absolute phase $\alpha $ has no physical
meaning. What can be measured is the phase difference between
superconductors. This means either staying in the mesoscopic
algebra of $\pi _c$ or comparing the inequivalent representations
$\pi _{\alpha}$ and $\pi_{\alpha'}$.
\end{enumerate}


\begin{thebibliography}{99}
\bibitem{F} P. Fendley, K. Schoutens, J. de Boer,
{\it Phys. Rev. Lett.} {\bf 90} (2003) 120402;\\
P. Fendley, B. Nienhuis, K. Schoutens, {\it J. Phys. A} {\bf 36}
(2003) 12399--12424.
\bibitem{Wit} E. Witten, {\it Nucl. Phys. B} {\bf 185} (1981) 513.
\bibitem{GVV} D. Goderis, A. Verbeure, P. Vets,
{\it Prob. Th. Rel. Fields}, {\bf 82} (1989) 527; {\it Commun.
Math. Phys.} {\bf 128} (1990) 533; {\it Il Nuovo Cim.} {\bf 106B}
(1991) 375.
\bibitem{HN} H. Narnhofer, {\it Found. Phys. Lett.} {\bf 17}
(2004)235--255.
\bibitem{PM} Ph.A. Martin, F. Rothen, {\it Many-Body Problems and Quantum Field Theory. An
Introduction}, Second Ed. (Springer Verlag, Berlin Heidelberg,
2004).

\bibitem{TW} W. Thirring, A. Wehrl, {\it Commun. Math. Phys.} {\bf
4} (1967) 303
\bibitem{VZ} A. Verbeure, V.A. Zagrebnov, {\it J. Stat. Phys.
}{\bf 69} (1992) 329
\bibitem {S} G. Sewell, {\sl Quantum Mechanics and its Emergent Macrophysics}
(Princeton Univ. Press, Princeton and Oxford, 2002).
\end{thebibliography}
\end{document}